\documentclass{iau-JDSS}
\usepackage{graphicx}

\title[Globular clusters and the Halo] 
{Globular cluster contributions to Galactic halo assembly}

\author[Sarah Martell]   
{Sarah L. Martell$^1$}

\affiliation{$^1$Australian Astronomical Observatory, North Ryde NSW 2122, Australia\break email: sarahmartell@aao.gov.au}

\pubyear{2012}
\volume{Volume 16}  
\pagerange{??}
\date{23 Oct. 2012 and in revised form ??}
\jname{Highlights of Astronomy, Volume 16}
\editors{Thierry Montmerle, ed.}
\begin{document}

\maketitle

\begin{abstract}
I discuss a search for red giant stars in the Galactic halo with light-element abundances similar to second-generation globular cluster stars, and discuss the implications of such a population for globular cluster formation models and the balance between \textit{in situ} star formation and accretion for the assembly of the Galactic halo.
\end{abstract}

\section{Introduction}\label{sec:intro}
We presently interpret the C-N, O-Na and Mg-Al abundance anticorrelations in globular cluster (GC) stars as a result of two-generation star formation and stellar-mode (i.e., supernova-free) chemical feedback (e.g., Carretta et al. 2010; D'Ercole et al. 2008). The present-day ratio of second- to first-generation stars in GCs is roughly 1:1, which leads to what is known as the ``mass budget problem'': there is simply not enough mass in stellar winds from the first generation to produce an equally massive second generation. Proposed solutions have included a top-heavy IMF for the first generation (Decressin et al. 2007) and a truncated mass function for the second generation (D'Ercole et al. 2008), but currently favored models require that the first generation was originally more massive (by a factor of 10-20) than it currently is. These massive GC formation models then require that the excess first-generation stars be preferentially removed from the cluster to reduce the ratio of second- to first-generation stars to its present-day level.

There are GCs that are currently losing stars to the halo field through extended tidal tails (e.g., Palomar 5, Odenkirchen et al. 2003; NGC 5466, Belokurov et al. 2006), and there is a theoretical expectation that many more GCs should have dissolved at earlier times as a result of tidal interactions with the Galaxy, internal 2-body interactions, and stellar evolution (Gnedin \& Ostriker 1997). If these globular clusters contained second-generation stars at the point of dissolution, then some fraction of halo field stars should carry the second-generation light-element abundance pattern.

\section{Globular cluster migrants in the halo field}\label{sec:data}
To look for halo field stars with second-generation abundances, Martell \& Grebel (2010) and Martell et al. (2011) searched the Sloan Digital Sky Survey (SDSS) SEGUE and SEGUE-2 low-resolution spectroscopic databases, respectively. Selecting red giants with $-1.8 \leq$ [Fe/H] $\leq -1.0$, reasonably well-determined stellar parameters and clean spectra, they identified a total of 2519 halo giants, 65 of which ($\sim 2.5\%$) appear to have second-generation carbon and nitrogen abudances. For these stars, the $3883\hbox{\AA}$ CN band is strong and the $4320\hbox{\AA}$ CH G-band is weak, relative to other field stars at similar metallicity and evolutionary phase. 


Ongoing work (Carollo et al., in prep.) is finding that these CN-strong field giants, presumably second-generation globular cluster stars that have been lost to the halo, have orbits and kinematics consistent with the inner halo population (Carollo et al. 2007; 2010). In that work, we are also finding that globular clusters with proper motion measurements available in the literature\footnote{Available at http://www.astro.yale.edu/dana/gc.html} have orbits and kinematics similar to the inner halo population, making them a reasonable potential source for {\it in situ} formation of the inner halo.


\section{Conclusions}\label{sec:concl}
The fraction of stars in the Galactic halo with light-element abundances similar to second-generation globular cluster stars is small, roughly $2.5\%$, but high-mass models for globular cluster formation require that they should be accompanied by several times as many first-generation stars, chemically indistinguishable from halo stars that formed outside GCs. This implies that globular clusters, as a major site of star formation 12 Gyr ago, are a significant contributor to Galactic halo assembly. 

\begin{acknowledgments}
SDSS-III is managed by the Astrophysical Research Consortium for the Participating Institutions of the SDSS-III Collaboration including the University of Arizona, the Brazilian Participation Group, Brookhaven National Laboratory, University of Cambridge, University of Florida, the French Participation Group, the German Participation Group, the Instituto de Astrofisica de Canarias, the Michigan State/Notre Dame/JINA Participation Group, Johns Hopkins University, Lawrence Berkeley National Laboratory, Max Planck Institute for Astrophysics, New Mexico State University, New York University, Ohio State University, Pennsylvania State University, University of Portsmouth, Princeton University, the Spanish Participation Group, University of Tokyo, University of Utah, Vanderbilt University, University of Virginia, University of Washington, and Yale University.
\end{acknowledgments}

\end{document}